\documentclass[aip, apl, amsmath, amssymb, reprint, superscriptaddress]{revtex4-2}
\usepackage{graphicx}
\usepackage{color}
\usepackage{siunitx}
\usepackage{bm}

\usepackage[
    colorlinks=true,
    linkcolor=blue,
    citecolor=blue,
    urlcolor=blue,
]{hyperref}

\newcommand{\Tstage}{T_{\mathrm{stage}}}

\newcommand{\Tc}{T_{\mathrm{c}}}
\newcommand{\BzS}{B_z^{\mathrm{meas}}}
\newcommand{\BzA}{B_z^{\mathrm{appl}}}
\newcommand{\Bcool}{B_z^{\mathrm{cooling}}}

\newcommand{\dth}{d_{\mathrm{th}}}
\newcommand{\Rnum}[1]{\uppercase\expandafter{\romannumeral #1\relax}}

\begin{document}

\title{Wide-field magnetic imaging of shielding-current-driven vortex rearrangement under local heating using diamond quantum sensors}

\author{Ryoei Ota}
\affiliation{Department of Physics, The University of Tokyo, Bunkyo-ku, Tokyo 113-0033, Japan\looseness=-1}
\author{Shunsuke Nishimura}
\affiliation{Department of Physics, The University of Tokyo, Bunkyo-ku, Tokyo 113-0033, Japan\looseness=-1}
\author{Koki Honda}
\affiliation{Department of Applied Physics, Tohoku University, Sendai, Miyagi 980-8579, Japan\looseness=-1}
\author{Takeyuki Tsuji}
\affiliation{International Center for Young Scientists, National Institute for Materials Science, Tsukuba, Ibaraki 305-0044, Japan}
\affiliation{Department of Electrical and Electronic Engineering, Institute of Science Tokyo, Meguro-ku, Tokyo 152-8552, Japan}
\author{Taro Yamashita}
\affiliation{Department of Applied Physics, Tohoku University, Sendai, Miyagi 980-8579, Japan\looseness=-1}
\author{Takayuki Iwasaki}
\affiliation{Department of Electrical and Electronic Engineering, Institute of Science Tokyo, Meguro-ku, Tokyo 152-8552, Japan}
\author{Mutsuko Hatano}
\affiliation{Department of Electrical and Electronic Engineering, Institute of Science Tokyo, Meguro-ku, Tokyo 152-8552, Japan}
\author{Kento Sasaki}
\affiliation{Department of Physics, The University of Tokyo, Bunkyo-ku, Tokyo 113-0033, Japan\looseness=-1}
\author{Kensuke Kobayashi}
\affiliation{Department of Physics, The University of Tokyo, Bunkyo-ku, Tokyo 113-0033, Japan\looseness=-1}

\begin{abstract}
Understanding and controlling vortex motion in superconductors are important both for suppressing dissipation in superconducting devices and for device applications that exploit vortices.
In this work, we quantitatively imaged the stray magnetic field distribution of vortices in an NbN thin film by wide-field magnetic imaging using a perfectly aligned diamond nitrogen-vacancy ensemble.
By continuously measuring while stepwise varying the applied magnetic field under local laser heating,
we captured a rearrangement of the vortex configuration in real space and in real time over more than \qty{100}{\min}.
The observed vortex rearrangement is consistent with a reduction of the pinning force due to local laser heating and with the Lorentz force exerted by shielding currents induced by the field variation.
These results provide insight into vortex dynamics and suggest potential applications, including vortex exclusion from sensitive regions of superconducting devices.
\end{abstract}

\maketitle

The motion of vortices in superconductors causes energy dissipation, making its suppression a critical issue in the design of superconducting devices.\cite{Song2009-wr, Yamashita2011-vj, Semenov2016-sb}
Vortex control is also important for exploiting vortices in devices, as indicated by studies suggesting their involvement in the detection mechanism of superconducting strip photon detectors (SSPDs)\cite{Bulaevskii2012-cu} and by proposed vortex-based device concepts.\cite{Wakamura2024-tm, Golod2015-gd, Lee1999-jb, Keren2023-xy}

To control vortex motion, it is essential to visualize vortex configurations in real space.
Imaging of vortex configurations has been achieved by various methods.\cite{Essmann1967-nz, Harada1992-yc, Goa2001-ps, Embon2017-oi}
Magnetic imaging using nitrogen-vacancy (NV) centers in diamond\cite{Rondin2012-yg, Rondin2014-vb, Steinert2010-km, Schirhagl2014-xj, Fu2020-mx}
has also been used to image the stray magnetic fields of vortices.\cite{Pelliccione2016-gc, Thiel2016-gg, Schlussel2018-hs, Acosta2019-od, Kapur2026-ha}
Using a perfectly aligned NV ensemble oriented normal to the surface as the sensor,\cite{Miyazaki2014-nx, Tahara2015-ob, Ozawa2017-uc, Ishiwata2017-jc, Tsuji2022-pn}
we previously achieved quantitative wide-field imaging of vortex stray-field distributions and have extended this approach to various systems\cite{Nishimura2023-my, Nishimura2025-xm, Shulga2025-ws}.
Here, we apply this method to visualize the stray magnetic field distribution of vortices in an NbN thin film, a material used in various superconducting devices, including SSPDs, owing to its excellent properties.\cite{Yamashita2013-ml, Takeuchi2024-gv, Watanabe2024-lu, Kim2024-ie}

\begin{figure}
    \centering
    \includegraphics[width=\linewidth]{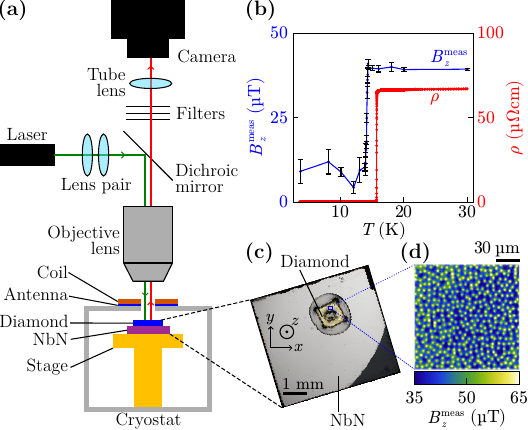}
    \caption{
    (a) Schematic of the experimental setup.
    (b)
    Red: Temperature dependence of the electrical resistivity $\rho$ of the NbN sample.
    Blue: Temperature dependence of the mean of the magnetic field $\BzS$ over a $\qty{55}{\micro\meter}\times\qty{55}{\micro\meter}$ region at the center of the field of view measured after cooling the sample to \qty{3.6}{\kelvin} in zero field, applying \qty{43}{\micro\tesla}, and then increasing the temperature. The temperature is that of the cryostat stage. Measurements were repeated 10 times at each temperature and the averages are plotted. Error bars represent 95\% confidence intervals.
    (c) Optical microscope image of the sample. 
    (d) Magnetic field distribution measured over a \qty{130}{\micro\meter} $\times$ \qty{130}{\micro\meter} field of view on the sample (blue square in panel (c)) after cooling the stage to \qty{3.6}{\kelvin} under an applied field of \qty{-47.4}{\micro\tesla}.
    }
    \label{fig:ExpSetup}
\end{figure}

\begin{figure*}
    \centering
    \includegraphics[width=\textwidth]{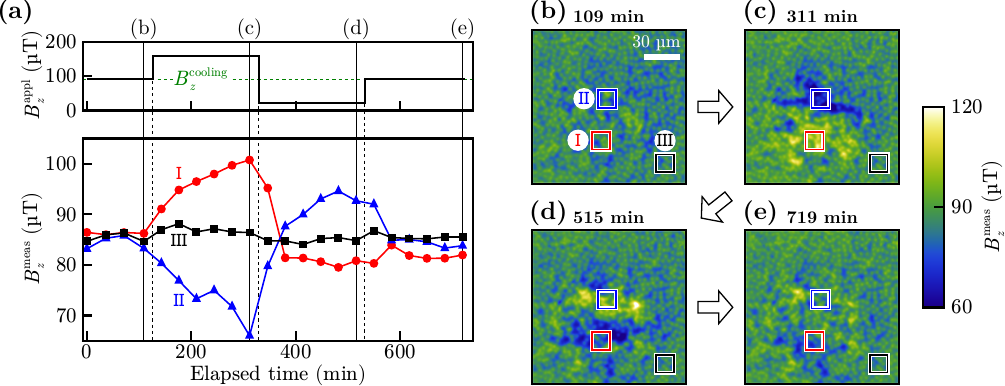}
    \caption{
        After cooling the sample under an applied field of $\Bcool=\qty{91}{\micro\tesla}$, the spatial distribution of the sample field $\BzS$ was continuously measured while the applied field $\BzA$ was changed stepwise at fixed stage temperature $\Tstage = \qty{11.5}{\kelvin} < \Tc'$. 
        (a) Top: Time dependence of applied field $\BzA$. Bottom: Time dependence of the spatially averaged value of measured field $\BzS$ in the three regions \Rnum{1}--\Rnum{3} indicated by squares in (b). Red circles, blue triangles, and black squares correspond to regions \Rnum{1}, \Rnum{2}, and \Rnum{3}, respectively.
        (b)--(e) Snapshots of the $\BzS$ distribution after holding the applied field at each step for a certain time.
    }
    \label{fig:VaryB_VtxMove}
\end{figure*}

The laser illumination used to excite NV centers can locally heat superconducting samples and perturb vortex configurations, as demonstrated in a previous study.\cite{Lillie2020-yo}
At the same time, optical control has attracted attention as a versatile route to generate, annihilate or manipulate vortices in superconductors.\cite{Plastovets2022-jy, Niedzielski2025-qw, Rochet2020-yj,Veshchunov2016-wq}

In this work, we captured, in real space and in real time, the evolution of the vortex configuration over more than \qty{100}{\min} after changes in the applied magnetic field in a region where pinning is weakened by local laser heating.
We show that this vortex motion is driven by the Lorentz force arising from shielding currents flowing in the sample, providing insight into the mechanism of vortex rearrangement.
Such vortex rearrangement may also be useful for excluding vortices from vortex-sensitive regions in superconducting devices, making the present result important as a proof of concept.

Figure~\ref{fig:ExpSetup}(a) shows a schematic of the experimental setup.
An NbN thin film with a diamond sensor attached was mounted on the stage of an optical cryostat (Montana Instruments Cryostation s50).
To obtain the magnetic field distribution by continuous-wave optically detected magnetic resonance (CW-ODMR),\cite{Rondin2014-vb, Nishimura2023-my} microwaves were applied to the diamond sample from an antenna placed directly above the cryostat,\cite{Sasaki2016-kq} while the sample was simultaneously illuminated with a 515-nm laser beam having an approximately Gaussian profile with a $1/e^2$ radius of about \qty{100}{\micro\meter}.
During both zero-field cooling and field cooling, the excitation laser was kept off to prevent laser-induced local heating from affecting the initial vortex distribution. 
Further details of the laser illumination conditions are provided in the supplementary material.
An out-of-plane magnetic field was applied to the sample using a coil placed directly above the cryostat.
The $z$ component of the applied field at the sensor position, $\BzA$, was determined from a calibration curve relating the coil current to the magnetic field measured beforehand using NV centers, with +$z$ defined as the out-of-plane upward direction.

The \qty{200}{nm}-thick NbN thin film was epitaxially grown on an $\mathrm{MgO}(100)$ substrate by reactive dc magnetron sputtering in an N$_2$/Ar atmosphere.
The background pressure was $2.0 \times 10^{-5}\,\mathrm{Pa}$, the Ar:N$_2$ flow-rate ratio was 100:8,
the deposition pressure was $2\,\mathrm{mTorr}$, and the sputtering current was $1.0\,\mathrm{A}$.
The temperature dependence of the electrical resistivity $\rho$ of the NbN sample, measured by the four-probe method using a Physical Property Measurement System (PPMS, Quantum Design), is shown by the red line in Fig.~\ref{fig:ExpSetup}(b). The sample exhibits a sharp superconducting transition at $\Tc = \qty{15.7}{\kelvin}$.
The diamond sensor was grown by chemical vapor deposition (CVD) on a $(111)$ Ib diamond substrate $(\qty{1}{mm} \times \qty{1}{mm} \times \qty{0.5}{mm})$, with NV centers incorporated within a $\qty{2.3}{\micro\meter}$-thick near-surface layer (NV-layer) and the NV axis controlled to be perfectly aligned along the out-of-plane direction.\cite{Miyazaki2014-nx, Tahara2015-ob, Ozawa2017-uc, Ishiwata2017-jc, Tsuji2022-pn}
The diamond sensor was attached to the NbN film with the NV-layer surface facing the film as shown in Fig. \ref{fig:ExpSetup}(c).

Using this setup, we measured the spatial distribution of the magnitude of the out-of-plane magnetic-field component, $\BzS$, just above the NbN sample in the \qty{130}{\micro\meter} $\times$ \qty{130}{\micro\meter} region indicated by the blue-box of Fig.~\ref{fig:ExpSetup}(c).
We first measured the temperature dependence of the diamagnetic response of NbN.
After cooling the sample to \qty{3.6}{\kelvin} in zero field and then applying \qty{43}{\micro\tesla},
we increased the temperature and measured the mean $\BzS$ over a $\qty{55}{\micro\meter}\times\qty{55}{\micro\meter}$ region at the center of the field of view; the result is shown by the blue line in Fig.~\ref{fig:ExpSetup}(b).
At low temperature, the measured field is sufficiently small, indicating a diamagnetic response.
At a stage temperature of $\Tstage = \qty{14.1}{\kelvin} \equiv \Tc'$, the measured field increases abruptly and the diamagnetic response disappears.
This indicates a transition of the sample from the superconducting to the normal state and is consistent with the temperature dependence of the electrical resistivity (red line in Fig.~\ref{fig:ExpSetup}(b)).
The difference between the transition temperatures, $\Delta T = \Tc - \Tc' = \qty{1.6}{\kelvin}$, is likely due to global heating of the sample by the laser and microwaves used for the magnetic measurements.

Next, we show a magnetic image obtained after field cooling.
Figure~\ref{fig:ExpSetup}(d) shows the magnetic field image obtained after cooling to \qty{3.6}{\kelvin} under an applied field of $\BzA = \qty{-47.4}{\micro\tesla}$.
In the \qty{130}{\micro\meter} $\times$ \qty{130}{\micro\meter} field of view, \num{399} local magnetic peaks were observed,
and the peak fields were nearly uniform, with a mean of \qty{62.6}{\micro\tesla} and a standard deviation of \qty{2.1}{\micro\tesla}.
Furthermore, the number of observed peaks agrees within \qty{3.1}{\percent} with the number estimated from the applied field,
$\qty{47.4}{\micro\tesla} \times (\qty{130}{\micro\meter})^2 / \Phi_0 = 387$,
assuming that each vortex carries a single flux quantum $\Phi_0 = \qty{2068}{\micro\tesla\cdot\micro\meter\squared}$.
This indicates that the observed peaks correspond to vortices carrying a single flux quantum.

We next investigated the motion of vortices.
After cooling the sample in a magnetic field of $\Bcool=\qty{91}{\micro\tesla}$, we repeatedly performed magnetic imaging while switching the applied field as shown in the top panel of Fig.~\ref{fig:VaryB_VtxMove}(a), with the stage temperature held at $\qty{11.5}{\kelvin}<\Tc'$.
The acquisition time for each image was about \qty{5}{\min}.
The bottom panel of Fig.~\ref{fig:VaryB_VtxMove}(a) shows, at intervals of approximately 35 min, the time evolution of the mean field in the three $\qty{15}{\micro\meter}\times\qty{15}{\micro\meter}$ regions \Rnum{1}--\Rnum{3} indicated in Fig.~\ref{fig:VaryB_VtxMove}(b).
Figures~\ref{fig:VaryB_VtxMove}(b)--(e) show snapshots of the field distribution after holding the applied field at each step for a certain time.
A movie obtained by concatenating all magnetic images is provided in the supplementary material.

Figure~\ref{fig:VaryB_VtxMove}(b) is a snapshot taken while the applied field was held at $\Bcool$.
No pronounced spatial inhomogeneity is observed in the magnetic field distribution.
While the applied field was kept at $\Bcool$, the mean field values in regions \Rnum{1}--\Rnum{3} changed very little, as shown in the bottom panel of Fig.~\ref{fig:VaryB_VtxMove}(a).
After the applied field was then increased to $\Bcool+\qty{68}{\micro\tesla}$, the field in region \Rnum{1} increased, whereas that in region \Rnum{2} decreased, yielding the field distribution shown in Fig.~\ref{fig:VaryB_VtxMove}(c).
This corresponds to vortex motion from top to bottom near the center of the field of view.
In contrast, the vortex configuration in the outer part of the field of view did not change.
Indeed, the mean field in region \Rnum{3} remained almost constant.
Next, after the applied field was decreased to $\Bcool - \qty{68}{\micro\tesla}$, the field in region \Rnum{1} decreased whereas that in region \Rnum{2} increased, giving the distribution shown in Fig.~\ref{fig:VaryB_VtxMove}(d).
This corresponds to vortex motion from bottom to top, opposite to that in Figs.~\ref{fig:VaryB_VtxMove}(b) and ~\ref{fig:VaryB_VtxMove}(c).
Finally, after the applied field was returned to the cooling value $\Bcool$, the field distribution became approximately uniform again (Fig.~\ref{fig:VaryB_VtxMove}(e)), and the fields in regions \Rnum{1}--\Rnum{3} recovered close to their initial values.
Throughout the continuous measurements, the total flux in the entire $\qty{130}{\micro\meter}\times\qty{130}{\micro\meter}$ field of view remained nearly constant, with a mean of $707\Phi_0$ and a standard deviation of $8.7\Phi_0$ ($\sim 1.2\%$), supporting the interpretation that the observed field changes reflect rearrangement of the vortex configuration within the field of view.
Additional measurements demonstrating that a vortex configuration displaced at elevated temperature can be stabilized by subsequent recooling are provided in the supplementary material.

\begin{figure}
    \centering
    \includegraphics[width=\linewidth]{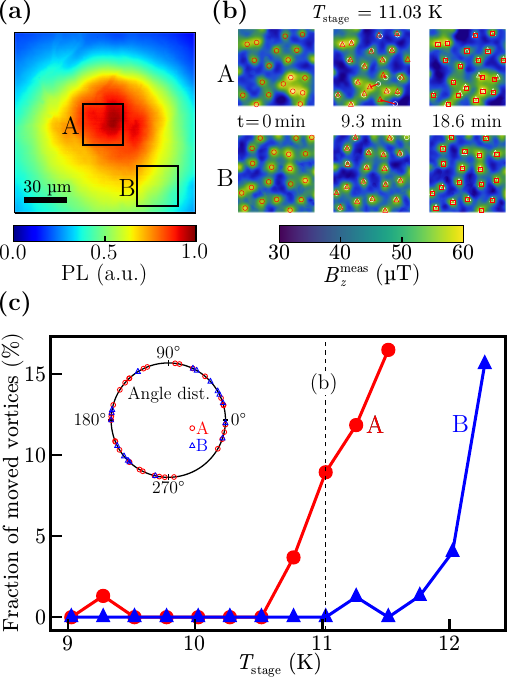}
    \caption{
        (a) Spatial distribution of the diamond photoluminescence (PL) intensity under laser illumination during the ODMR measurement.
        The black squares indicate the $\qty{30}{\micro\meter}\times\qty{30}{\micro\meter}$ regions A and B used for the analysis.
        (b) Time evolution of the magnetic field maps $\BzS$ in regions A and B ($t = 0$, 9.3, and \qty{18.6}{\min}) obtained after cooling the sample in a magnetic field of \qty{46}{\micro\tesla} and holding it at $\Tstage = \qty{11.03}{\kelvin}$.
        Red markers: Peak positions extracted from the field maps ($t = \qty{0}{\min}$, circles; $t = \qty{9.3}{\min}$, triangles; $t = \qty{18.6}{\min}$, squares).
        White markers: Peak positions at the preceding time.
        Red arrows: Pairs of nearest peaks between successive frames separated by more than the threshold $\dth = \qty{1}{\micro\meter}$.
        (c) Stage-temperature dependence of the fraction of moved vortices.
        Red circles and blue triangles correspond to regions A and B, respectively.
        Inset: Scatter plot of vortex-motion angles for all temperatures.
    }
    \label{fig:depairing_Temp-RegionDep}
\end{figure}

During the continuous magnetic imaging, changes in the vortex configuration were observed near the center of the field of view, whereas the configuration in the outer region remained unchanged.
This suggests that pinning is weaker near the center of the field of view.
In the CW-ODMR measurements, laser light of approximately \qty{20}{\milli\watt} illuminates a region with a radius of about \qty{100}{\micro\meter},
and we infer that the spatial profile of the laser intensity causes local heating of the sample.
Figure~\ref{fig:depairing_Temp-RegionDep}(a) shows the spatial distribution of the NV photoluminescence (PL) intensity in this measurement.
Because the PL intensity is proportional to the incident laser intensity, the laser intensity is higher near the center of the field of view.
The local temperature is therefore expected to be higher there, leading to weaker pinning.\cite{Allen1989-vu, Goldstein1989-jd}

To examine the temperature dependence of vortex depinning, we kept the magnetic field constant after field cooling and consecutively imaged the vortex stray-field distribution five times at each temperature in regions A and B (Fig.~\ref{fig:depairing_Temp-RegionDep}(a)), tracking the motion of individual vortices.
The acquisition time for one image was approximately 9 min.
Figure~\ref{fig:depairing_Temp-RegionDep}(b) shows an example at a stage temperature of \qty{11.03}{\kelvin}.
Here, the peak positions extracted from the field maps are indicated by red markers, and nearby pairs of peaks between successive frames were identified.
Pairs separated by more than the threshold $\dth = \qty{1}{\micro\meter}$ were regarded as vortex hopping events and connected by arrows.
Several hopping events were detected in region A, whereas none were detected in region B.

Figure~\ref{fig:depairing_Temp-RegionDep}(c) shows the mean fraction of vortices that moved between frames at each temperature.
An increase in the depinning fraction is observed above a stage temperature of \qty{10.75}{\kelvin} in region A and above \qty{11.75}{\kelvin} in region B.
Thus, increasing temperature clearly weakens the pinning.
Furthermore, because depinning appears in region A at a stage temperature approximately \qty{1}{\kelvin} lower than in region B,
region A is inferred to be approximately $\Delta T_{\mathrm{AB}}=\qty{1}{\kelvin}$ hotter than region B.
This temperature difference explains why the vortex configuration changed only near the center of the field of view in Fig.~\ref{fig:VaryB_VtxMove}.
Although the difference in mobility could arise from spatial inhomogeneity of the pinning landscape, an additional experiment performed with a weaker and broader laser-illumination profile did not show such a pronounced difference between the two regions, as described in the supplementary material.
This comparison supports the interpretation that the spatially nonuniform vortex mobility in Figs.~\ref{fig:VaryB_VtxMove} and ~\ref{fig:depairing_Temp-RegionDep}(c) is primarily caused by the laser-induced temperature distribution.

The inset of Fig.~\ref{fig:depairing_Temp-RegionDep}(c) shows a scatter plot of the angles of all detected vortex motions, which are approximately isotropic.
It indicates the motions are thermally activated or driven by random local perturbations, such as fluctuations in laser intensity.
Another possible contribution from laser heating is a driving force arising from the temperature gradient itself.
However, such a force would be expected to induce directional vortex motion, whereas the observed motion is nearly isotropic. Therefore, the temperature-gradient force is not considered to be the main contribution.
This interpretation is also supported by Fig.~\ref{fig:VaryB_VtxMove}, in which vortices moved in both directions with respect to the temperature gradient.
\begin{figure}
    \centering
    \includegraphics[width=\linewidth]{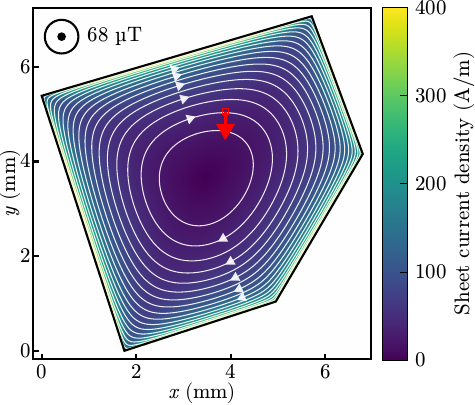}
    \caption{
        Distribution of the sheet current density $K$ calculated under the assumption of complete shielding of an increment of \qty{68}{\micro\tesla} applied along the positive $z$ direction of the sample, corresponding to the field increase in Figs.~\ref{fig:VaryB_VtxMove}(b) and ~\ref{fig:VaryB_VtxMove}(c).
        The black pentagon shows the NbN sample outline, matched to the actual dimensions in Fig.~\ref{fig:ExpSetup}(c).
        The white lines are current streamlines (contours of the stream function), and the current flows along them in the directions indicated by the white arrows.
        The red box (\qty{130}{\micro\meter} $\times$ \qty{130}{\micro\meter}) marks the ODMR measurement area,
        and the red arrow indicates the direction of the Lorentz force acting on a vortex with flux along the +z axis estimated from the local sheet current density at the center of the measurement area.
    }
    \label{fig:CalcShieldingCurrent}
\end{figure}

Next, we discuss the relationship between the applied-field changes and the vortex rearrangement in Fig.~\ref{fig:VaryB_VtxMove}.
When the applied magnetic field is changed, a current flows in the sample so as to shield the field change, and the vortices are expected to move under the Lorentz force exerted by this shielding current.
We therefore calculated the shielding-current distribution for the field increment of \qty{68}{\micro\tesla} in Figs.~\ref{fig:VaryB_VtxMove}(b) and ~\ref{fig:VaryB_VtxMove}(c).\cite{Prigozhin2018-qh}
Here, we assumed complete shielding of the field change.
Since the Pearl length $\Lambda$, which characterizes the lateral extent of a vortex, is clearly much smaller than the sample width $w \sim \qty{5}{mm}$ ($\Lambda \ll w$), the assumption of complete shielding is justified.\cite{Babaei-Brojeny2003-zl}
Indeed, during the applied-field changes in Fig.~\ref{fig:VaryB_VtxMove}(a), the field in the outer part of the field of view and the total flux in the entire field of view changed very little, indicating that complete shielding is realized.

The calculated shielding-current distribution is shown in Fig.~\ref{fig:CalcShieldingCurrent}, with the current flowing along the white streamlines.
At the magnetic imaging area indicated by the red box, the sheet current density is $K = |\bm{K}| = \qty{37}{A\per\meter}$, corresponding to a current density of $j = K/d = \qty{1.8e-2}{\mega\ampere\per\centi\metre\squared}$ where $d=\qty{200}{nm}$ is the NbN film thickness. 
From the local sheet current density in the measurement area, we calculate the Lorentz force $\bm{K} \times (\Phi_0 \bm{e}_z)$, acting on a vortex with flux along the $+z$ axis. The force is directed as indicated by the red arrow in Fig.~\ref{fig:CalcShieldingCurrent} and its magnitude is \qty{7.6e-2}{\pico\newton}.
This force direction agrees with the observed direction of vortex motion when the applied field is increased [Figs.~\ref{fig:VaryB_VtxMove}(b) and ~\ref{fig:VaryB_VtxMove}(c)].
Converting this value to a pinning force per unit length gives a value about one order of magnitude smaller than that measured for an NbN thin film at \qty{12}{\kelvin} by magnetic force microscopy (MFM) in a previous study,\cite{Shapoval2011-nl}
but the difference is not inconsistent with the weaker pinning expected at the higher local temperature,\cite{Allen1989-vu, Goldstein1989-jd} roughly estimated as
$\Tstage + \Delta T + \Delta T_\mathrm{AB} =\qty{14.1}{\kelvin}$.
We note that this local temperature is only an estimate and was not directly measured.
For a more quantitative discussion of the local sample temperature, an independently controlled heater and less invasive measurement schemes that minimize laser- and microwave-induced heating would be useful~\cite{Lillie2020-yo, Sengottuvel2025-eh, Dhungel2026-ks, Kapur2026-ha}.
When the applied field is decreased, the shielding current reverses direction and so does the force acting on the vortices, explaining the direction of vortex motion in Figs.~\ref{fig:VaryB_VtxMove}(c) and ~\ref{fig:VaryB_VtxMove}(d).
When the applied field is returned to the cooling value, the shielding current decreases and the vortex-vortex repulsion relaxes the spatial bias (Figs.~\ref{fig:VaryB_VtxMove}(d) and ~\ref{fig:VaryB_VtxMove}(e)).

In summary, using a perfectly aligned NV ensemble, we quantitatively imaged the stray magnetic field distribution of vortices in an NbN thin film.
Under local heating and global magnetic-field variation, we further captured changes in the vortex configuration in real space and in real time.
Exploiting the advantages of wide-field imaging, we statistically analyzed the temperature dependence of vortex motion and theoretically calculated the driving force generated by the shielding current under field variation, enabling a consistent explanation of the observed vortex motion.

This study provides important insight into the driving forces acting on vortices and the temperature dependence of pinning.
The ability to create a selected vortex-mobile region well away from the film edge and to apply a tunable, directional force driven by shielding currents offers a useful means of analyzing depinning characteristics in the film interior.
Moreover, by combining local heating with magnetic-field variation, the present method can collectively move vortices in a certain direction within a selected region, which may enable applications such as vortex exclusion from sensitive regions of superconducting devices.

\bigskip
See the supplementary material for details of the laser illumination conditions,
a movie showing the full time sequence of the magnetic-field images corresponding to Fig.~\ref{fig:VaryB_VtxMove},
results demonstrating the stabilization of the displaced vortex configuration by recooling,
and vortex reconfiguration under weaker and broader laser irradiation.

\bigskip
This work is partially supported by 
JST, CREST Grant Number JPMJCR23I2, Japan;
JST SPRING, Grant Number JPMJSP2108;
JSPS Grants-in-Aid for Scientific Research (Nos. JP26H02007, JP23K25800, JP24K21194, JP25K00934, JP24KJ0657, and JP25H01248); 
MEXT Quantum Leap Flagship Program (MEXT Q-LEAP) Grant Number JP-MXS0118067395; 
``Advanced Research Infrastructure for Materials and Nanotechnology in Japan (ARIM)'' of the Ministry of Education, Culture, Sports, Science and Technology (MEXT), Proposal Number JPMXP1225UT1155; 
the Cooperative Research Project of RIEC, Tohoku University;
New Challenge Research hosted by JSR Corporation via JSR-UTokyo Collaboration Hub, CURIE. 

\section*{AUTHOR DECLARATIONS}
\subsection*{Conflict of Interest}
The authors have no conflicts to disclose.
\subsection*{Author Contributions}
\textbf{Ryoei Ota}: Conceptualization (lead); Data curation (lead); Formal analysis (lead); Investigation (lead); Methodology (equal); Software (lead); Validation (lead); Visualization (lead); Writing – original draft (lead); Writing – review \& editing (equal).  \textbf{Shunsuke Nishimura}: Conceptualization (equal); Investigation (supporting); Methodology (equal); Software (equal); Writing – review \& editing (equal). \textbf{Koki Honda}: Data curation (equal); Investigation (equal); Resources (lead); Writing – review \& editing (equal). \textbf{Takeyuki Tsuji}: Resources (lead); Writing – review \& editing (equal). \textbf{Taro Yamashita}: Conceptualization (equal); Supervision (equal); Writing – review \& editing (equal). \textbf{Takayuki Iwasaki}: Supervision (equal); Writing – review \& editing (equal). \textbf{Mutsuko Hatano}: Supervision (equal); Writing – review \& editing (equal). \textbf{Kento Sasaki}: Conceptualization (equal); Funding acquisition (equal); Project administration (equal); Supervision (equal); Writing – review \& editing (lead). \textbf{Kensuke Kobayashi}: Conceptualization (equal); Funding acquisition (lead); Project administration (lead); Supervision (lead); Writing – review \& editing (lead).
\section*{DATA AVAILABILITY}
The data that support the findings of this study are available from the corresponding authors upon reasonable request.

\bibliography{refs}

\end{document}


\title{Supplemental Information for:\\
Wide-field magnetic imaging of shielding-current-driven vortex rearrangement under local heating using diamond quantum sensors
}

\author{Ryoei Ota}
\affiliation{Department of Physics, The University of Tokyo, Bunkyo-ku, Tokyo 113-0033, Japan\looseness=-1}
\author{Shunsuke Nishimura}
\affiliation{Department of Physics, The University of Tokyo, Bunkyo-ku, Tokyo 113-0033, Japan\looseness=-1}
\author{Koki Honda}
\affiliation{Department of Applied Physics, Tohoku University, Sendai, Miyagi 980-8579, Japan\looseness=-1}
\author{Takeyuki Tsuji}
\affiliation{International Center for Young Scientists, National Institute for Materials Science, Tsukuba, Ibaraki 305-0044, Japan}
\affiliation{Department of Electrical and Electronic Engineering, Institute of Science Tokyo, Meguro-ku, Tokyo 152-8552, Japan}
\author{Taro Yamashita}
\affiliation{Department of Applied Physics, Tohoku University, Sendai, Miyagi 980-8579, Japan\looseness=-1}
\author{Takayuki Iwasaki}
\affiliation{Department of Electrical and Electronic Engineering, Institute of Science Tokyo, Meguro-ku, Tokyo 152-8552, Japan}
\author{Mutsuko Hatano}
\affiliation{Department of Electrical and Electronic Engineering, Institute of Science Tokyo, Meguro-ku, Tokyo 152-8552, Japan}
\author{Kento Sasaki}
\affiliation{Department of Physics, The University of Tokyo, Bunkyo-ku, Tokyo 113-0033, Japan\looseness=-1}
\author{Kensuke Kobayashi}
\affiliation{Department of Physics, The University of Tokyo, Bunkyo-ku, Tokyo 113-0033, Japan\looseness=-1}

\maketitle

\section{Details of the laser illumination conditions}
A 515-nm laser (LBX-515, Oxxius) was used for optical excitation.
The output was coupled into a multimode fiber, and a De-Speckler (Newport) was used to average out speckle in the illumination profile.
For all cooling procedures, including both zero-field cooling and field cooling, the excitation laser was kept off to avoid laser-induced local heating during the cooling process.
Unless otherwise stated, the standard laser-illumination condition described below was used for all measurements in the main text and in this Supplementary Material, except for Sec.~\ref{uniform_laser}.
The laser power after the objective lens was \qty{20}{\milli\watt}, as measured with an optical power meter (Thorlabs).
The beam size at the sample plane was adjusted using a pair of lenses.
The illumination profile at the sample plane, estimated from the spatial distribution of the NV photoluminescence under laser illumination [Fig.~3(a) of the main text], was approximately Gaussian, with a \(1/e^2\) radius of about \qty{100}{\micro\meter}.

In Sec.~\ref{uniform_laser}, we used a weakened and broadened laser-illumination condition for comparison to examine the effect of the spatial distribution of laser-induced heating on vortex motion. The corresponding NV photoluminescence (PL) distribution is shown in Fig.~\ref{fig:Fig_LargeLaserSpot}(a), together with that obtained under the standard laser-illumination condition. Under this condition, the PL variation across the field of view is smaller than the standard condition, indicating a more uniform temperature distribution.

\section{Caption for Movie S1}
The full time sequence of the magnetic-field images corresponding to Fig.~2 in the main text.

Alt Text: Time-sequence video of magnetic-field images acquired while the temperature is held constant and the magnetic field is changed stepwise. A gradual change in the magnetic-field distribution is observed.

\section{Stabilizing the displaced vortex configuration by recooling}

\begin{figure}
    \centering
    \includegraphics[width=\linewidth]{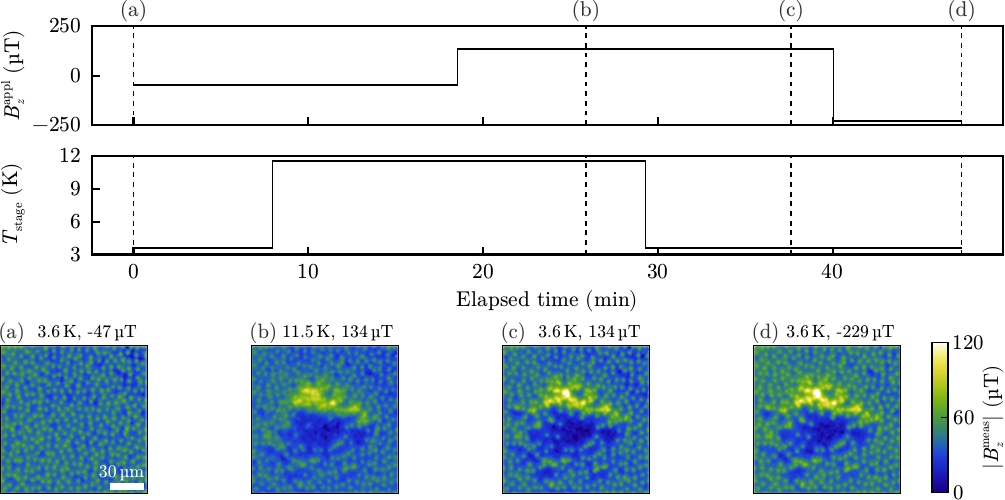}
    \caption{
    Top: Time sequences of the applied field $B^{\mathrm{appl}}_{z}$ and stage temperature $T_{\mathrm{stage}}$ after field cooling under \qty{-47}{\micro\tesla}.
    Bottom: Corresponding magnetic-field images at the four time points indicated by the dashed lines in the top panels. The measurement position and field of view are the same as those in the main text (blue box in Fig.~1(c) of the main text). \\
    Alt Text:
    Stepwise changes in applied field and stage temperature are correlated with four magnetic-field images. A localized stray-field feature appears during heating under the changed field and persists after recooling and after a subsequent field change, showing that the displaced vortex configuration is retained.
    }
    \label{fig:Demo_VtxMove}
\end{figure}

In Fig.~2 of the main text, we showed vortex displacement driven by shielding currents induced by changes in the applied field at elevated temperature.
In Figure~\ref{fig:Demo_VtxMove}, we further show that recooling stabilizes the displaced vortex configuration.

After cooling the sample to \qty{3.6}{\kelvin} under an applied field of \qty{-47}{\micro\tesla}, we varied the applied field and the stage temperature as shown in the upper panels of Fig.~\ref{fig:Demo_VtxMove}.
Fig.~\ref{fig:Demo_VtxMove}(a) shows the magnetic-field image obtained after field cooling, with vortices visible in the field of view.
Note that the vortices carry flux along the $-z$ direction because the field during cooling was directed along $-z$.

When the stage temperature was raised to \qty{11.5}{\kelvin} and the applied field was changed to \qty{134}{\micro\tesla}, the magnetic-field distribution shifted toward the upper part of the image, as shown in Fig.~\ref{fig:Demo_VtxMove}(b).
This direction of motion is consistent with the Lorentz force expected for vortices carrying flux along the $-z$ direction, that is, opposite to that in Fig.~4 of the main text.

Next, the temperature was lowered again to \qty{3.6}{\kelvin}, and the resultant configuration is shown in Fig.~\ref{fig:Demo_VtxMove}(c).
This configuration remained unchanged even when the applied field was changed to \qty{-229}{\micro\tesla}, which would induce a Lorentz force in the opposite direction, as shown in Fig.~\ref{fig:Demo_VtxMove}(d).
This result indicates that the pinning force was stronger than both the mutual repulsion between vortices and the Lorentz force arising from the shielding current.

These results demonstrate the potential of combining local heating with changes in the applied field to move vortices away from vortex-sensitive regions in superconducting devices and then stabilize the resulting configuration by recooling.

\section{Vortex reconfiguration under more uniform laser-induced heating conditions} \label{uniform_laser}
\begin{figure}
    \centering
    \includegraphics[width=\linewidth]{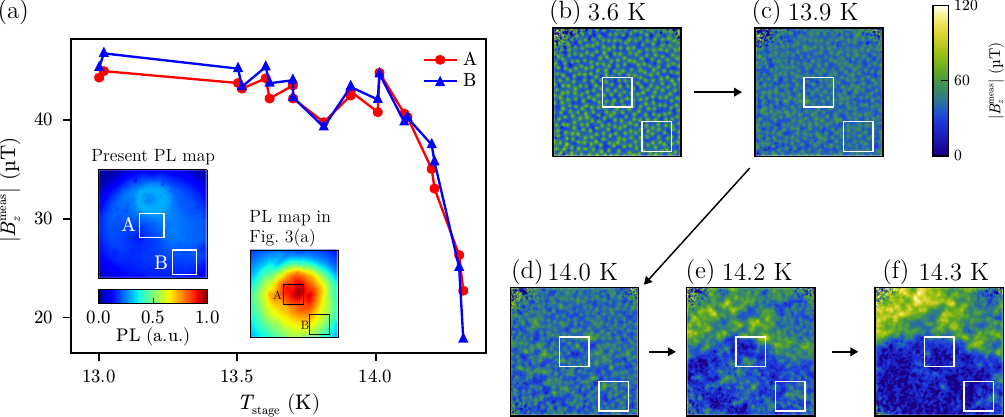}
    \caption{
    Temperature evolution of the magnetic field distribution under more spatially uniform laser irradiation. The left inset in (a) shows the present NV-center PL distribution under the broadened laser-irradiation condition, whereas the right inset shows the reference PL distribution under the standard laser-irradiation condition used in the main text, corresponding to Fig.~3(a). The same color scale is used for both PL maps, and both maps are normalized by the maximum PL intensity in Fig.~3(a). (a) Stage-temperature dependence of the average magnetic field, $|B_z^{\mathrm{meas}}|$, in regions A and B. (b)--(f) Snapshots of $|B_z^{\mathrm{meas}}|$ at $T_{\mathrm{stage}} = 3.6$, 13.9, 14.0, 14.2, and $14.3~\unit{\kelvin}$. \\
    Alt Text: 
    Figure S2 shows vortex reconfiguration under broadened laser irradiation. Panel (a) plots the average measured magnetic field magnitude in regions A and B as a function of stage temperature; the two curves follow nearly the same temperature dependence and decrease rapidly above about 14.2 K. The left inset shows a weak and nearly uniform NV photoluminescence distribution for the present broadened laser condition, while the right inset shows a more strongly peaked reference photoluminescence distribution for the standard laser condition used in the main text. Panels (b)–(f) show magnetic-field snapshots at 3.6, 13.9, 14.0, 14.2, and 14.3 K. The vortex configuration remains nearly unchanged up to 13.9 K, starts to change over the field of view at 14.0 K, and is substantially rearranged at 14.2–14.3 K, with vortices expelled toward the upper part of the field of view.
    }
    \label{fig:Fig_LargeLaserSpot}
\end{figure}

In the main text, we observed that vortices move more readily in the central region of the field of view, where the laser intensity is high, and attributed this behavior to laser-induced local heating.
However, vortices are trapped at pinning sites, and the in-plane distribution of the pinning strength may also contribute to their mobility. Therefore, to examine whether laser-induced local heating is indeed the dominant factor, we analyzed the reconfiguration of vortices under weaker and broader laser irradiation in the same field of view.
The photoluminescence (PL) map of the NV centers under laser excitation in this experiment, together with the PL map in Fig. 3(a) for comparison, is shown in the inset of Fig. \ref{fig:Fig_LargeLaserSpot}(a). Note that the color scale is common to both maps and is normalized by the maximum value of the PL distribution in Fig. 3(a). Compared with the PL distribution in Fig.~3(a), the present PL intensity is lower and its spatial variation is smaller.
The absolute difference between the PL intensities at A and B corresponds to only 4\% of the maximum PL intensity in Fig.~3(a).

After cooling the sample to 3.6 K under a magnetic field of \qty{-47}{\micro\tesla} with the laser turned off, the magnetic field was changed to \qty{134}{\micro\tesla}. The sample was then illuminated by the laser to measure magnetic field maps. Subsequently, the temperature was increased and the magnetic field maps at each temperature were analyzed. The temperature dependence of the average magnetic field in regions A and B is shown in Fig.\ref{fig:Fig_LargeLaserSpot}(a).
In contrast to the results of Fig. 2 of the main text, which show clearly different behavior between central and peripheral regions of the field of view, the average magnetic fields in regions A and B exhibited nearly identical temperature dependences.
The temperature evolution of the magnetic field distribution is shown in Fig. \ref{fig:Fig_LargeLaserSpot}(b)--(f). At 13.9 K, the vortex configuration remained unchanged compared with that at 3.6 K. At 14.0 K, the vortex configuration began to change over the entire field of view, and at 14.2–14.3 K, vortices were expelled toward the upper part of the field of view. This behavior is in contrast to the result shown in Fig. 2, where the vortex configuration changed only near the center of the field of view. These results support the interpretation that the preferential vortex motion observed near the center of the field of view in the main text is primarily governed by the laser-heating distribution rather than by a spatial variation in the pinning landscape.

